\documentclass[aps,prl,twocolumn,floatfix,superscriptaddress]{revtex4-2}
\usepackage{graphicx,amsmath,amsfonts,mathrsfs,bbold,dsfont}
\begin{document}

\title{Magnetically induced polarization in centrosymmetric bonds}

\author{Igor Solovyev}
\email{SOLOVYEV.Igor@nims.go.jp}
\affiliation{National Institute for Materials Science, MANA, 1-1 Namiki, Tsukuba,
Ibaraki 305-0044, Japan}
\affiliation{Department of Theoretical Physics and Applied Mathematics, Ural Federal University,
Mira str. 19, 620002 Ekaterinburg, Russia}
\affiliation{Institute of Metal Physics, S. Kovalevskaya str. 18, 620108 Ekaterinburg, Russia}
\author{Ryota Ono}
\affiliation{Graduate School of Science and Engineering, Chiba University, 1-33 Yayoi-cho, Inage-ku, Chiba-shi 265-8522, Japan}
\author{Sergey Nikolaev}
\affiliation{National Institute for Materials Science, MANA, 1-1 Namiki, Tsukuba,
Ibaraki 305-0044, Japan}
\affiliation{Tokyo Tech World Research Hub Initiative (WRHI), Institute of Innovative Research, Tokyo Institute of Technology, 4259 Nagatsuta, Midori-Ku, Yokohama, Kanagawa, 226-8503, Japan}
\affiliation{Laboratory for Materials and Structures, Tokyo Institute of Technology, 4259 Nagatsuta, Midori-Ku, Yokohama, Kanagawa, 226-8503, Japan}

\date{\today}

\date{\today}
\begin{abstract}
We reveal the microscopic origin of electric polarization $\vec{P}$ induced by noncollinear magnetic order. We show that in Mott insulators, such $\vec{P}$ is given by all possible combinations of position operators $\hat{\vec{r}}_{ij} = (\vec{r}_{ij}^{\, 0},\vec{\boldsymbol{r}}_{ij}^{\phantom{0}})$ and transfer integrals $\hat{t}_{ij} = (t_{ij}^{0},\boldsymbol{t}_{ij}^{\phantom{0}})$ in the bonds, where $\vec{r}_{ij}^{\, 0}$ and $t_{ij}^{0}$ are spin-independent contributions in the basis of Kramers doublet states, while $\vec{\boldsymbol{r}}_{ij}^{\phantom{0}}$ and $\boldsymbol{t}_{ij}^{\phantom{0}}$ stem solely from the spin-orbit interaction. Among them, the combination $t_{ij}^{0} \vec{\boldsymbol{r}}_{ij}^{\phantom{0}}$, which couples to the spin current, remains finite in the centrosymmetric bonds, thus yielding finite $\vec{P}$ in the case of noncollinear arrangement of spins. The form of the magnetoelectric coupling, which is controlled by $\vec{\boldsymbol{r}}_{ij}^{\phantom{0}}$, appears to be rich and is not limited to the phenomenological law $\vec{P} \sim \boldsymbol{\epsilon}_{ij} \times [\boldsymbol{e}_{i} \times \boldsymbol{e}_{j}]$ with $\boldsymbol{\epsilon}_{ij}$ being the bond vector connecting the spins $\boldsymbol{e}_{i}$ and $\boldsymbol{e}_{j}$. Using  density-functional theory, we illustrate how the proposed mechanism work in the spiral magnets CuCl$_2$, CuBr$_2$, CuO, and $\alpha$-Li$_2$IrO$_3$, providing consistent explanation to available experimental data.
\end{abstract}

%\keywords{}
\maketitle

\par \emph{Introduction}. In order to make a material ferroelectric, it is essential to break the inversion symmetry. Canonically, this implied some crystallographic instability towards polar atomic displacements~\cite{FE}. Nevertheless, there is a very special class of materials, called multiferroics, where inversion symmetry can be broken by magnetic means in an otherwise perfect crystallographically centrosymmetric lattice~\cite{MF}. The microscopic origin of  multiferroicity is the fundamental physical problem and its practical realization is an important step towards mutual control of electric polarization and magnetism in novel electronic devices.

\par There can be various scenarios of the magnetic inversion symmetry breaking. In certain multiferroic materials,  the inversion symmetry is microscopically broken by local distortions, so that individual bonds can be formally associated with some polarization vectors. When arranged in an antiferroelectric manner, these bonds result in zero net polarization. Nevertheless, if some of them become inequivalent owing to complex magnetic order, the perfect cancellation of polarization vectors does not occur and the system becomes ferroelectric.

\par However, what if the bond itself is centrosymmetric? Can it become electrically polarized by magnetic means? An affirmative answer to these questions was given by Katsura, Nagaosa, and Balatsky (KNB)~\cite{KNB}, who considered a very special microscopic model and argued that the noncollinear alignment of spins can induce the polarization $\vec{P} \sim \boldsymbol{\epsilon}_{ij} \times [\boldsymbol{e}_{i} \times \boldsymbol{e}_{j}]$, which lies in the plane of spins $\boldsymbol{e}_{i}$ and $\boldsymbol{e}_{j}$ and is perpendicular to the bond vector $\boldsymbol{\epsilon}_{ij}$. This finding was supported by phenomenological considerations~\cite{Mostovoy} and the proposed mechanism was called the ``spin-current mechanism'', which is widely used for the analysis of magnetoelectric (ME) coupling in spiral magnets~\cite{TokuraSeki}, typically in combination with two other mechanisms -- the ``exchange striction''~\cite{Sergienko,Malashevich} and ``spin-dependent $p$-$d$ hybridization''~\cite{TokuraSekiNagaosa}.

\par Nevertheless, the analysis remains largely phenomenological. First, the properties of all spiral multiferroics are usually discussed from the viewpoint of the spin-current model~\cite{Seki_CuCl2,Zhao_CuBr2,Kimura_TbMnO3,MnWO4,Kimura,LiCu2O2}. However, there are only few materials, such as CuCl$_2$~\cite{Seki_CuCl2} and CuBr$_2$~\cite{Zhao_CuBr2}, consisting solely of the centrosymmetric bonds. In other materials, the situation is not so straightforward~\cite{Kimura_TbMnO3,MnWO4,Kimura,LiCu2O2}: as the symmetry is low, the bonds are not necessarily centrosymmetric, thus allowing for alternative explanations~\cite{gavs2,alternative,PRB2019}. Then, if KNB model fails to explain the properties of spiral magnets, it typically causes some confusion with identifying the problem and choosing a suitable alternative~\cite{PRB2020}. Although density-functional theory (DFT) provides a powerful tool for calculating the polarization~\cite{FE_theory,WannierRevModPhys}, the formal mapping of DFT results on a specifically selected model~\cite{MnI2_Xiang} does not shed light on microscopic mechanisms underlying this model.

\par In this Letter, we formulate a transparent microscopic theory of electric polarization induced by noncollinear magnetic order. We explicitly show how and why the noncollinear arrangement of spins gives rise to electric polarization even in centrosymmetric bonds. We relate the ME coupling to fundamental symmetry properties of the position operator and transfer integrals in the basis of Kramers states and argue that the paradigm of the spin-current induced polarization appears to be much richer and goes beyond the phenomenological low $\vec{P} \sim \boldsymbol{\epsilon}_{ij} \times [\boldsymbol{e}_{i} \times \boldsymbol{e}_{j}]$~\cite{PRB2019,PRB2020,MnI2_Xiang}. We evaluate all relevant parameters on the basis of DFT and show how they are manifested in the properties of real spiral magnets.

\par \emph{Basic theory}. The simplest toy model, which captures the physics, is the 1-orbital Hubbard model with spin-orbit interaction (SOI),
\noindent
\begin{equation}
\hat{\cal{H}}  =  \sum\limits_{ij} \sum_{\sigma \sigma'}
t_{ij}^{\sigma \sigma'}
\hat{c}^{\dagger}_{i \sigma}
\hat{c}^{\phantom{\dagger}}_{j \sigma'} +
U \sum\limits_{i}
\hat{n}_{i \uparrow} \hat{n}_{i \downarrow},
\label{eqn.ManyBodyH}
\end{equation}
\noindent where $\hat{c}^{\dagger}_{i \sigma}$ ($\hat{c}^{\phantom{\dagger}}_{i \sigma}$) creates (annihilates) a hole with pseudospin $\sigma$$=$ $+$ or $-$ at site $i$, $\hat{n}_{i \sigma} = \hat{c}^{\dagger}_{i \sigma} \hat{c}^{\phantom{\dagger}}_{i \sigma}$, $\hat{t}_{ij} = [ t_{ij}^{\sigma \sigma'} ]$ are the transfer integrals, and  $U$ is the on-site Coulomb repulsion. The Wannier functions, $|i \sigma \rangle = \hat{c}^{\dagger}_{i \sigma}|0 \rangle$, can be chosen as the Kramers pairs, transforming under the time reversal as  $\hat{T}|i$$\mp \rangle = \pm |i$$\pm \rangle$. All model parameters are derived from DFT~\cite{DFT2Model}, in the subspace of Wannier states, which are primarily responsible for the magnetism~\cite{WannierRevModPhys,wannier90}. The Coulomb $U$ is evaluated within constrained random-phase approximation~\cite{cRPA}.

\par Since the polarization in metals is screened by free electrons, the multiferroicity is the property of insulating state. Then, the problem can be solved in the spirit of superexchange (SE) theory, by treating $\hat{t}_{ij}$ as a perturbation~\cite{Anderson}. Let $| \alpha_{i}^{o} \rangle = \cos \frac{\theta_{i}}{2}|i$$+ \rangle - \sin \frac{\theta_{i}}{2} \, e^{-i \phi_{i}} |i$$- \rangle$ be the occupied orbital in the limit $\hat{t}_{ij}$$=$$0$, where $\theta_{i}$ and $\phi_{i}$ specify the direction $\boldsymbol{e}_{i} = (\cos \phi_{i} \sin \theta_{i}, \sin \phi_{i} \sin \theta_{i}, \cos \theta_{i})$ of spin, and $| \alpha_{i}^{u} \rangle = \sin \frac{\theta_{i}}{2} \, e^{ i \phi_{i}} |i$$+ \rangle + \cos \frac{\theta_{i}}{2} |i$$- \rangle$ is the unoccupied orbital. To the first order in $\frac{\hat{t}_{ij}}{U}$, $| \alpha_{i}^{o} \rangle$ will transform to the Wannier function $| w_{i} \rangle = | \alpha_{i}^{o} \rangle + \sum_{j} | \alpha_{i \to j}^{o} \rangle$, acquiring the tails
\noindent
\begin{equation}
| \alpha_{i \to j}^{o} \rangle = -\frac{1}{U} | \alpha_{j}^{u} \rangle \langle \alpha_{j}^{u} | \hat{t}_{ji} | \alpha_{i}^{o} \rangle
\label{eq:wtails}
\end{equation}
\noindent on surrounding sites $j$. Considering expectation values of the kinetic energy $\langle w_{i} | \hat{t}_{ij} | w_{i} \rangle$, one can readily formulate the spin model
\noindent
\begin{equation}
\mathcal{E} = \sum\limits_{\langle i j\rangle} \left( - J_{ij}\boldsymbol{e}_{i} \cdot \boldsymbol{e}_{j} + \boldsymbol{D}_{ij} \cdot [\boldsymbol{e}_{i}\times\boldsymbol{e}_{j}] + \boldsymbol{e}_{i} \cdot \tensor{\Gamma}_{ij} \boldsymbol{e}_{j} \right),
\label{eq:spinmodel}
\end{equation}
\noindent describing the energy change in terms of isotropic ($J_{ij}$), Dzyaloshinskii-Moriya (DM, $\boldsymbol{D}_{ij}$), and symmetric anisotropic ($\tensor{\Gamma}_{ij}$) interactions. A similar model can be formulated for  polarization~\cite{Moriya_pol}:
\noindent
\begin{equation}
\vec{P} = \sum\limits_{\langle i j\rangle} \left( \vec{\mathsf{P}}_{ij} \, \boldsymbol{e}_{i} \cdot \boldsymbol{e}_{j} + \vec{\boldsymbol{\mathcal{P}}}_{ij} \cdot [ \boldsymbol{e}_{i} \times \boldsymbol{e}_{j} ] + \boldsymbol{e}_{i} \cdot \vec{\boldsymbol{\Pi}}_{ij} \boldsymbol{e}_{j} \right),
\label{eq:p1o}
\end{equation}
\noindent in terms of the vector $\vec{\mathsf{P}}_{ij} \equiv [\mathsf{P}_{ij}^{v}]$, rank-2 tensor $\vec{\boldsymbol{\mathcal{P}}}_{ij} \equiv [ \mathcal{P}_{ij}^{v,c} ]$, and rank-3 tensor $\vec{\boldsymbol{\Pi}}_{ij} \equiv [{\Pi}_{ij}^{v,ab}]$~\cite{footnote4}. The model parameters can be obtained from matrix elements of the position operator in the framework of general theory for polarization in periodic systems~\cite{FE_theory,WannierRevModPhys}:
\noindent
\begin{equation}
\vec{P}=-\frac{e}{V}\sum\limits_{i} \langle w_{i} | \vec{r} \, | w_{i} \rangle
\label{eq:elpol}
\end{equation}
\noindent (where $V$ is the volume and $-$$e$ is the electron charge). Then, the only matrix elements that contribute to the magnetic dependence of $\vec{P}$ are of the type $\langle \alpha_{i}^{o} | \vec{r} \, | \alpha_{i \to j}^{o} \rangle$. Other contributions, such as $\langle \alpha_{i \to j}^{o} | \vec{r} \, | \alpha_{i \to j}^{o} \rangle$ or single-ion anisotropy of $\vec{P}$, vanish in the 1-orbital model~\cite{PRB2019,PRB2020}.

\par The $2$$\times$$2$ matrices $\hat{t}_{ij}$ and $\hat{\vec{r}}_{ij}$ can be decomposed in terms of the unity $\hat{\mathbb{1}}$ and the vector $\hat{\boldsymbol{\sigma}} = (\hat{\sigma}_{x},\hat{\sigma}_{y},\hat{\sigma}_{z})$ of Pauli matrices as $\hat{t}_{ij} = t_{ij}^{0} \hat{\mathbb{1}} + i \boldsymbol{t}_{ij}^{\phantom{0}} \hat{\boldsymbol{\sigma}}$ and $\hat{\vec{r}}_{ij} = \vec{r}_{ij}^{\, 0} \hat{\mathbb{1}} + i \vec{\boldsymbol{r}}_{ij}^{\phantom{0}} \hat{\boldsymbol{\sigma}}$ with the real coefficients $(t_{ij}^{0},\boldsymbol{t}_{ij}^{\phantom{0}})$ and $(\vec{r}_{ij}^{\, 0},\vec{\boldsymbol{r}}_{ij}^{\phantom{0}})$. Furthermore, the hermiticity yields $t_{ji}^{0} = t_{ij}^{0}$, $\boldsymbol{t}_{ji} = -\boldsymbol{t}_{ij}$, $\vec{r}_{ji}^{\, 0} = \vec{r}_{ij}^{\, 0}$, and $\vec{\boldsymbol{r}}_{ji} = -\vec{\boldsymbol{r}}_{ij}$~\cite{footnote3}. The corresponding spin model parameters are summarized in Table~\ref{tab:1orb}~\cite{SM}.
\noindent
\begin{table}[b]
\caption{Isotropic ($J_{ij}$ and $\vec{\mathsf{P}}_{ij}$), antisymmetric ($\boldsymbol{D}_{ij}$ and $\vec{\boldsymbol{\mathcal{P}}}_{ij}$), and anisotropic symmetric ($\boldsymbol{\Gamma}_{ij} = \boldsymbol{\Gamma}_{ij}'-\frac{1}{2}{\rm Tr}\boldsymbol{\Gamma}_{ij}'$ and $\vec{\boldsymbol{\Pi}}_{ij} = \vec{\boldsymbol{\Pi}}_{ij}' - \frac{1}{2}{\rm Tr}\vec{\boldsymbol{\Pi}}_{ij}'$) parameters of exchange interactions and polarization. $\otimes$ denotes the direct product. }
\label{tab:1orb}
\begin{ruledtabular}
\begin{tabular}{cc}
       exchange                      & polarization \\
\hline
$J_{ij} = -\frac{(t_{ij}^{0})^2}{U}$    & $\vec{\mathsf{P}}_{ij} = -\frac{2e}{V} \frac{\vec{r}_{ij}^{\, 0} t_{ij}^{0}}{U}$         \\
$\boldsymbol{D}_{ij} = \frac{2t_{ij}^{0}\boldsymbol{t}_{ij}^{\phantom{0}}}{U}$    & $\vec{\boldsymbol{\mathcal{P}}}_{ij} = -\frac{2e}{V} \frac{ \vec{r}_{ij}^{\, 0} \boldsymbol{t}_{ij}^{\phantom{0}} + t_{ij}^{0} \vec{\boldsymbol{r}}_{ij}^{\phantom{0}}}{U}$         \\
$\boldsymbol{\Gamma}_{ij}' = \frac{2 \boldsymbol{t}_{ij} \otimes \boldsymbol{t}_{ij} }{U}$ & $\vec{\boldsymbol{\Pi}}_{ij}' = -\frac{e}{V} \frac{\vec{\boldsymbol{r}}_{ij} \otimes \boldsymbol{t}_{ij} +  \boldsymbol{t}_{ij} \otimes \vec{\boldsymbol{r}}_{ij} }{U}$
\end{tabular}
\end{ruledtabular}
\end{table}

\par A very special case is when two sites $i$ and $j$ are connected by spacial inversion. Since $\hat{t}_{ij}$ is a scalar, but $\hat{\vec{r}}_{ij}$ is a (true) vector, inversion symmetry requires that $\hat{t}_{ji} = \hat{t}_{ij}$, but $\hat{\vec{r}}_{ji} = -\hat{\vec{r}}_{ij}$. In combination with hermiticity, we have $\boldsymbol{t}_{ij} = 0$ and $\vec{r}_{ij}^{\, 0} = 0$. Then, the DM interaction $\boldsymbol{D}_{ij}$ vanishes, which is the general property of centrosymmetric (\texttt{c}) bonds, and so does  $\boldsymbol{\Gamma}_{ij}$, being not independent in the 1-orbital model~\cite{hiddens}. Thus, the only interaction in the \texttt{c}-bonds will be $J_{ij}$. The behavior of electric polarization is different: $\vec{\mathsf{P}}_{ij}=0$ and $\vec{\boldsymbol{\Pi}}_{ij}=0$, while  $\vec{\boldsymbol{\mathcal{P}}}_{ij} = -\frac{2e}{V} \frac{ t_{ij}^{0} \vec{\boldsymbol{r}}_{ij}^{\phantom{0}}}{U} \equiv \vec{\boldsymbol{\mathcal{C}}}_{ij}$ can persist even in the \texttt{c}-bond (in agreement with symmetry arguments~\cite{MnI2_Xiang}), and the corresponding polarization, $\vec{\boldsymbol{\mathcal{C}}}_{ij}  \cdot [ \boldsymbol{e}_{i} \times \boldsymbol{e}_{j} ]$, is induced solely by the noncollinear arrangement of spins.

\par In the noncentrosymmetric (\texttt{nc}) bonds, $\boldsymbol{t}_{ij}$ and $\vec{r}_{ij}^{\, 0}$ are finite, resulting in  nonzero $\boldsymbol{D}_{ij}$, $\boldsymbol{\Gamma}_{ij}$, $\vec{\mathsf{P}}_{ij}$, and $\vec{\boldsymbol{\Pi}}_{ij}$. Moreover, $\vec{\boldsymbol{\mathcal{P}}}_{ij}$ acquires an additional asymmetric contribution $\vec{\boldsymbol{\mathcal{A}}}_{ij} = -\frac{2e}{V} \frac{ \vec{r}_{ij}^{\, 0} \boldsymbol{t}_{ij}^{\phantom{0}} }{U}$. $\vec{\mathsf{P}}_{ij}$ and $\vec{\boldsymbol{\mathcal{A}}}_{ij}$ can be viewed as a regular polarization of the \texttt{nc}-bond ($-\frac{e \vec{r}_{ij}^{\, 0}}{V}$), which is additionally modulated by the spin texture. If this texture results from the competition of $\boldsymbol{D}_{ij}$ and $J_{ij}$, a similar competition takes place between $\vec{\boldsymbol{\mathcal{A}}}_{ij}$ and $\vec{\mathsf{P}}_{ij}$, as it depends on the same ratio $\frac{\boldsymbol{t}_{ij}^{\phantom{0}}}{t_{ij}^{0}}$. Therefore, there will be a \emph{cancellation} of contributions associated with $\vec{\boldsymbol{\mathcal{A}}}_{ij}$ and $\vec{\mathsf{P}}_{ij}$~\cite{PRB2020}.

\par The 1-orbital model is subjected to hidden symmetries, that allow to fully eliminate $\boldsymbol{t}_{ij}^{\phantom{0}}$ by rotating the spins at the sites $i$ and $j$~\cite{hiddens}. In such local coordinate frame, the bond is solely described by the transfer integral $\tilde{t}_{ij}^{0} = \sqrt{ (t_{ij}^{0})^{2} + \boldsymbol{t}_{ij}^{\phantom{0}} \cdot \boldsymbol{t}_{ij}^{\phantom{0}} }$, exchange interactions -- by $J_{ij}$, and the electric polarization -- by the competition of $\vec{\mathsf{P}}_{ij}$ and $\vec{\boldsymbol{\mathcal{C}}}_{ij}$ (with $\tilde{t}_{ij}^{0}$ instead of $t_{ij}^{0}$). Using $| \alpha^{u}_{j} \rangle \langle \alpha^{u}_{j} | = \hat{\mathbb{1}} - | \alpha^{o}_{j} \rangle \langle \alpha^{o}_{j} |$ in Eq.~(\ref{eq:wtails}), each term in the spin model can be further expressed via the expectation value of some quantity in the ground state. Since $[ \boldsymbol{e}_{i} \times \boldsymbol{e}_{j} ]$ is related to the spin current $\hat{\boldsymbol{j}}_{ij}^{s}=\frac{i\tilde{t}_{ij}^{0}}{2}\big(\hat{\tilde{c}}^{\dagger}_{i\sigma}\hat{\boldsymbol{\sigma}}_{\sigma\sigma'}\hat{\tilde{c}}^{\phantom{\dagger}}_{j\sigma'}-\hat{\tilde{c}}^{\dagger}_{j\sigma}\hat{\boldsymbol{\sigma}}_{\sigma\sigma'}\hat{\tilde{c}}^{\phantom{\dagger}}_{i\sigma'} \big)$~\cite{BrunoDugaev}, one can find that $\vec{P} \sim \vec{\boldsymbol{\mathcal{C}}}_{ij} \cdot \langle \boldsymbol{j}^{S}_{ij} \rangle$~\cite{SM}, in analogy with the DM interactions~\cite{Kikuchi}.

\par The $3$$\times$$3$ tensor $\vec{\boldsymbol{\mathcal{C}}}_{ij} = [ \mathcal{C}_{ij}^{v,c} ]$ can be generally decomposed into symmetric and antisymmetric parts, $\vec{\boldsymbol{\mathcal{C}}}_{ij} = \vec{\boldsymbol{\mathcal{C}}}_{ij}^{S} + \vec{\boldsymbol{\mathcal{C}}}_{ij}^{A}$.  The latter is expressed as  $\varepsilon_{vca} \pi_{ij}^{a}$, in terms of the vector $\boldsymbol{\pi}_{ij} = [\pi_{ij}^{a}]$ and Levi-Civita symbol $\varepsilon_{vca}$. The corresponding $\vec{P} = \boldsymbol{\pi}_{ij} \times [ \boldsymbol{e}_{i} \times \boldsymbol{e}_{j} ] $ reminds of the KNB expression~\cite{KNB,Mostovoy}. Nevertheless, $\boldsymbol{\pi}_{ij}$ is not necessarily parallel to $\boldsymbol{\epsilon}_{ij}$, and $\vec{\boldsymbol{\mathcal{C}}}$ can include $\vec{\boldsymbol{\mathcal{C}}}_{ij}^{S}$. Thus, the proposed spin-current theory is not limited to the conventional law $\vec{P} \sim \boldsymbol{\epsilon}_{ij} \times [\boldsymbol{e}_{i} \times \boldsymbol{e}_{j}]$ and includes other interesting options. Below, we consider examples of how it works for different symmetries.

\par \emph{Relativistic $j$$=$$\frac{1}{2}$ manifold of $t_{2g}$ states}. In the cubic environment, the sixfold degenerate $t_{2g}$ states are split by SOI into fourfold degenerate $\Gamma_{8}$ states and the Kramers pair of $\Gamma_{7}$ states $| + \rangle = \frac{1}{\sqrt{3}} ( | xy$$\downarrow$$\rangle - | yz$$\uparrow$$\rangle + i | zx$$\uparrow$$\rangle )$ and $| - \rangle = \frac{1}{\sqrt{3}} ( | xy$$\uparrow$$\rangle + | yz$$\downarrow$$\rangle + i | zx$$\downarrow$$\rangle )$. The latter can be viewed as the effective $j=\frac{1}{2}$ pseudospin states, which play a key role in the physics of spin-orbit Mott insulators realized in $5d$ Ir oxides~\cite{Kim2008}. Since the KNB expression was derived assuming this symmetry of states~\cite{KNB}, we start our analysis with this example and consider a perfect bond obeying the $C_{\infty}^{z}$ symmetry along $z$. It is straightforward to show that the only nonzero elements of $\vec{\boldsymbol{r}}_{ij} = [ r_{ij}^{v,c} ]$ will be $r_{ij}^{y,x} = - r_{ij}^{x,y} = \frac{1}{3} \, ( \langle xy_{i} | y | zx_{j} \rangle - \langle zx_{i} | y | xy_{j} \rangle )$~\cite{SM}. The \emph{antisymmetric} tensor can be presented as $\varepsilon_{abz} \pi_{ij}^{z}$, where $ab=$ $xy$ or $yx$, and the vector $\boldsymbol{\pi}_{ij} = (0,0,\pi_{ij}^{z})$ is indeed parallel to the bond. Thus, we do recover the KNB expression $\vec{P} \sim \boldsymbol{\epsilon}_{ij} \times [\boldsymbol{e}_{i} \times \boldsymbol{e}_{j}]$~\cite{KNB}. Nevertheless, this form of $\vec{P}$ is the consequence of the \emph{particular symmetry of undistorted $t_{2g}$ states}. Other symmetries may yield different $\vec{P}$.

\par \emph{$z^{2}$ states with SOI}. The simplest example, illustrating this idea is the $z^{2}$ states, which are stabilized by the crystal field and mixed with the $yz$ and $zx$ states by SOI. The Kramers states are ~\cite{Takayama}: $| + \rangle \propto | z^{2}$$\downarrow$$\rangle + \xi ( i| yz$$\uparrow$$\rangle -| zx$$\uparrow$$\rangle ) $ and $| - \rangle \propto | z^{2}$$\uparrow$$\rangle + \xi ( i | yz$$\downarrow$$\rangle + | zx$$\downarrow$$\rangle )$ ($\xi$ being the ratio of SOI to the crystal field splitting), and the non-vanishing elements $r_{ij}^{v,c}$ are: $r_{ij}^{y,x} = r_{ij}^{x,y} \propto \langle z^{2}_{i} | x | zx_{j} \rangle - \langle zx_{i} | x | z^{2}_{j} \rangle$~\cite{SM}. Thus, the tensor $[ r_{ij}^{v,c} ]$ is \emph{symmetric}, meaning that for two noncollinear spins $\boldsymbol{e}_{1,2} = (\sin \theta \cos \varphi,\sin \theta \sin \varphi,\pm \cos \theta)$, the polarization $\vec{P} \sim (\cos \varphi,-\sin \varphi,0)$ is still perpendicular to the bond, but does not necessarily lie in the spin plane.

\par
\begin{figure}[b]
\begin{center}
\includegraphics[width=0.47\textwidth]{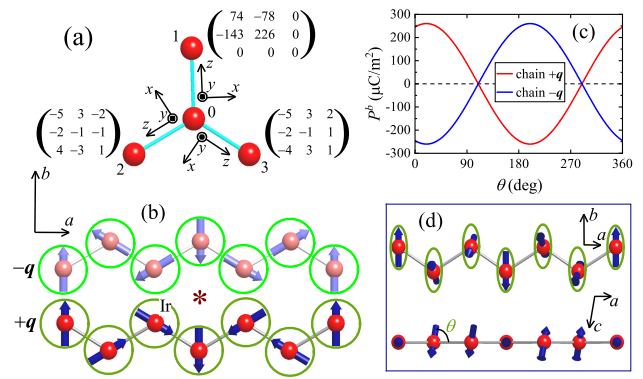}
\end{center}
\caption{
(a) Tensor $\vec{\boldsymbol{\mathcal{C}}}_{ij}$ in the nearest \texttt{c}-bonds of $\alpha$-Li$_2$IrO$_3$ (in $\mu$C/m$^2$), in the local coordinate frames denoted by $x$, $y$ and $z$. (b) Double-$\boldsymbol{q}$ magnetic structure realized in $\alpha$-Li$_2$IrO$_3$. Inversion center is denoted by $*$. (c) Magnetically induced polarization in the chains with $+\boldsymbol{q}$ and $-\boldsymbol{q}$ depending on the rotation of the spin-spiral plane about the monoclinic $b$ axis, as explained in (d) in the planes $ab$ and $bc$. $\theta$ is the angle formed by the spin-spiral plane and the monoclinic $a$ axis.}
\label{fig.Li2IrO3}
\end{figure}

 \emph{$\alpha$-Li$_2$IrO$_3$}. As the first realistic example, it is instructive to consider the spin-orbit Mott insulator $\alpha$-Li$_2$IrO$_3$, which attracted a great deal of attention as a possible material for realizing the Kitaev spin liquid state~\cite{Chaloupka}. In this monoclinic compound (space group $C2/m$), there are two types of nearest \texttt{c}-bonds: the strongest one ($01$ in Fig.~\ref{fig.Li2IrO3}a) along the monoclinic $b$ axis, which is transformed to itself by the twofold rotation about $b$; and two rotationally non-invariant weak bonds ($02$ and $03$), which are connected by the twofold rotation.
\noindent

\par The minimal model was constructed for the magnetic $j$$=$$\frac{1}{2}$ bands~\cite{SM}. In the local coordinate frame, where the $z$ axis is along the bond, the tensor $\vec{\boldsymbol{\mathcal{C}}}_{01}$ has only $x$ and $y$ components (see Fig.~\ref{fig.Li2IrO3}a). Hence, the polarization is perpendicular to the bond and, considering only the antisymmetric part of $\vec{\boldsymbol{\mathcal{C}}}_{01}$, we would indeed obtain $\vec{P} \sim \boldsymbol{\epsilon}_{01} \times [\boldsymbol{e}_{0} \times \boldsymbol{e}_{1}]$~\cite{KNB,Mostovoy}. However, besides the antisymmetric contribution, $\vec{\boldsymbol{\mathcal{C}}}_{01}$ clearly shows a strong symmetric one, which is expected for this type of  symmetry~\cite{footnote1}. In the low-symmetry bonds $02$ and $03$, $\vec{\boldsymbol{\mathcal{C}}}_{ij}$ is even more complex: all elements are finite and inequivalent, so that the polarization can be perpendicular as well as parallel to the bond.

\par Since $\alpha$-Li$_2$IrO$_3$ is a noncollinear magnet with $\boldsymbol{q} \approx (\frac{1}{3},0,0)$~\cite{aLi2IrO3}, it is interesting to ask whether it can become multiferroic. The magnetic texture can be viewed as the double-$\boldsymbol{q}$ spiral, propagating in alternating zigzag chains with $+\boldsymbol{q}$ and $-\boldsymbol{q}$ (Fig.~\ref{fig.Li2IrO3}b). The spiral order induces an appreciable polarization in each of the chains (Fig.~\ref{fig.Li2IrO3}c). However, the chains are connected by the inversion operation, the contributions with $+\boldsymbol{q}$ and $-\boldsymbol{q}$ cancel each other, and $\alpha$-Li$_2$IrO$_3$ remains antiferroelectric.

\par
\begin{figure}[b]
\begin{center}
\includegraphics[width=0.47\textwidth]{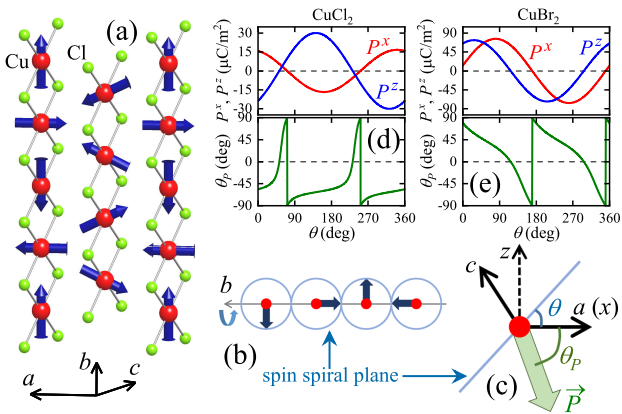}
\end{center}
\caption{
(a) Crystal and magnetic structure of CrCl$_2$ in the $ab$ plane. (b) Rotations of the spin spiral plane about the axis $b$. (c) Angles $\theta$ and $\theta_{P} = \tan^{-1} \frac{P^{z}}{P^{x}}$ specifying the spiral plane and electric polarization, respectively. (d) and (e) Angle dependence of electric polarization in CuCl$_2$ and CuBr$_2$.}
\label{fig.CuCl2}
\end{figure}

 \emph{Copper dihalides}. CuCl$_2$ and CuBr$_2$ form the chain-like monoclinic structure (space group $C2/m$), where all Cu-Cu bonds are centrosymmetric. Below the N\'eel temperature ($T_{\rm N}= 24$ and $74$ K for CuCl$_2$ and CuBr$_2$, respectively), they develop a cycloidal magnetic order with $\boldsymbol{q} \approx (1,\frac{1}{4},\frac{1}{2})$ (Fig.~\ref{fig.CuCl2}a), which coincides with the onset of ferroelectricity~\cite{Seki_CuCl2,Zhao_CuBr2}.
\noindent

\par The minimal model was constructed for the magnetic ``Cu $x^2$-$y^2$'' bands~\cite{SM}. The obtained parameters $J_{ij}$ nicely reproduce the experimental cycloidal order~\cite{footnote2}. Polarization induced by the cycloidal order is related to $\vec{\boldsymbol{\mathcal{C}}}_{ij}$. Since twofold rotations about $b$ transform the chains to themselves, the symmetry properties of $\vec{\boldsymbol{\mathcal{C}}}_{ij}$ in these chains are similar to $\vec{\boldsymbol{\mathcal{C}}}_{01}$ in  $\alpha$-Li$_2$IrO$_3$ (Fig.~\ref{fig.Li2IrO3}a) and the polarization is expected to be perpendicular to $b$. In the nearest bonds $\pm b$, the antisymmetric part of $\vec{\boldsymbol{\mathcal{C}}}_{\pm b}$ is given by $\boldsymbol{\pi}_{\pm b} = (0,\pm 0.5,0)$ and $(0,\pm 8,0)$ $\mu$C/m$^2$ for CuCl$_2$ and CuBr$_2$, respectively, which is too small to account for the total $\vec{P}$, and a substantially larger contribution stems from the symmetric part. The behavior of $\vec{P}$ (after summation over all bonds) is summarized in Fig.~\ref{fig.CuCl2}d and e. For CuCl$_2$, we note a good agreement with the experiment~\cite{Seki_CuCl2}, including deviation of $\vec{P}$ from the spin-spiral plane by the angle $\theta + \theta_{P}$. Similar behavior is expected for CuBr$_2$ with somewhat larger $T_{\rm N}$ and $\vec{P}$ due to stronger SOI and hybridization mediated by Br $4p$ states.

\par
\begin{figure}[t]
\begin{center}
\includegraphics[width=0.47\textwidth]{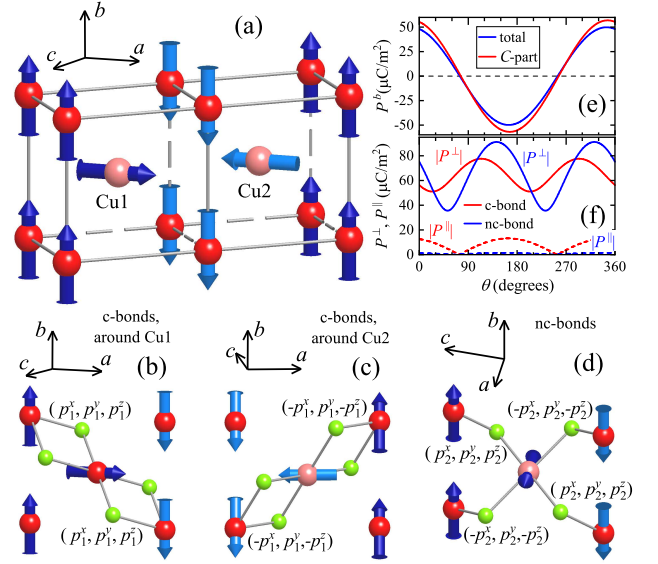}
\end{center}
\caption{
(a) Spiral magnetic order in CuO. (b)-(c) Nearest \texttt{c}- and (d) \texttt{nc}-bonds in the monoclinic planes $ab$ and $bc$ with the vectors of induced polarization. (e) Electric polarization (total and centrosymmetric part described by $\vec{\boldsymbol{\mathcal{C}}}_{ij}$) as the function of angle $\theta$ formed by the spins in the $ac$ plane with the axis $a$. Here, $x$ and $y$ are chosen along $a$ and $b$, respectively, and $z$ is perpendicular to $a$ and $b$. (f) Electric polarization perpendicular ($P^{\perp}$) and parallel ($P^{\parallel}$) to the bond, calculated for the nearest \texttt{c}- and \texttt{nc}-bonds as a function of $\theta$.}
\label{fig.CuO}
\end{figure}

 \emph{Cupric oxide}. CuO with the centrosymmetric monoclinic structure (space group $C2/c$) has attracted much attention as a simple binary material that becomes multiferroic at exceptionally high temperature~\cite{Kimura}. Ferroelectricity is driven by a spiral magnetic order with $\boldsymbol{q} \approx (\frac{1}{2},0,-\frac{1}{2})$ (see Fig.~\ref{fig.CuO}a) emerging between $213$ and $230$ K~\cite{CuOexp} (and is expected even at room temperature under hydrostatic pressure~\cite{CuORT}).
\noindent

\par Construction of the minimal model is similar to Cu dihalides~\cite{SM}. Although the crystal structure of CuO includes both \texttt{c}- and \texttt{nc}-bonds (Fig.~\ref{fig.CuO}), the parameters associated with the latter can be largely eliminated~\cite{footnote5}, while $\vec{\boldsymbol{\mathcal{C}}}_{ij}$ is responsible for 90 \% of $\vec{P}$ (Fig.~\ref{fig.CuO}e).

\par Since the symmetry is low, the form of $\vec{\boldsymbol{\mathcal{C}}}_{ij}$ is complex~\cite{SM}, and the polarization vectors in individual bonds are specified by all three projections, as shown in Fig.~\ref{fig.CuO}. Notably, there can be appreciable components along the bonds (Fig.~\ref{fig.CuO}f). Nevertheless, due to the twofold rotations about $y$ ($b$), only the $y$ component of $\vec{P}$ will survive after summation over all equivalent bonds. This is an important point where the symmetry comes into play: the experimental polarization is induced along the $y$ axis not because of the phenomenological rule $\boldsymbol{q} \times [ \boldsymbol{e}_{i} \times \boldsymbol{e}_{j} ] $~\cite{KNB,Mostovoy}. Quite the contrary, it is a consequence of the particular $C2/c$ symmetry of CuO. Total $P^{b} \approx 55$ $\mu$C/m$^2$ is comparable with the experimental $P^{b} \sim 150$ $\mu$C/m$^2$~\cite{Kimura,CuO_Cc}.

\par \emph{Conclusion}. We have presented a toy theory revealing the fundamental origin of the magnetic inversion symmetry breaking in centrosymmetric systems. Due to intrinsic symmetries of the transfer integrals and position operator in the basis of Kramers states, the combination $t_{ij}^{0} \vec{\boldsymbol{r}}_{ij}^{\phantom{0}}$ remains finite, yielding finite polarization for noncollinear spins. This polarization depends on the symmetry of Kramers states, providing new alternatives beyond phenomenological spin-current model. The abilities of proposed theory are illustrated on spiral magnets CuCl$_2$, CuBr$_2$, CuO, and $\alpha$-Li$_2$IrO$_3$.

\par \emph{Acknowledgement}. We are grateful to Anonymous Referee of Ref.~\cite{PRB2020} for drawing our attention to intersite matrix elements of $\vec{r}$. I.S. was supported by program AAAA-A18-118020190095-4 (Quantum).


\begin{thebibliography}{99}
\bibitem{footnote4}
The vector symbol and index $v$ are reserved for the direction of polarization, while the bold symbols and indices after the comma stand for the coupling to the spin part.
\bibitem{footnote1}
If the $\Gamma_{7}$ $t_{2g}$ states are distorted so that the $| xy \rangle$, $| yz \rangle$, and $| zx \rangle$ orbitals contribute to $| \pm \rangle$ with unequal weights, the tensor $\vec{\boldsymbol{r}}_{ij}$ is allowed to have simultaneously antisymmetric and symmetric parts~\cite{SM}.
\bibitem{footnote2}
There are four main interactions separated by monoclinic translations $b$, $2b$, $3b$, and $c$: $J_{b}=-$$0.3$ ($-$$0.4$), $J_{2b}=-$$3.0$ ($-$$3.9$), $J_{3b}=-0.1$ ($-$$0.2$), and $J_{c}=-$$0.7$ ($-$$1.4$) meV for CuCl$_2$ (CuBr$_2$), which are responsible for the spin-spiral order in the chains (along $b$) and the AFM coupling between the chains (along $c$). Corresponding $T_{\rm N}$, evaluated in the random phase approximation~\cite{SM}, is $32$ ($56$) K for CuCl$_2$ (CuBr$_2$).
\bibitem{footnote5}
By properly choosing local coordinate frames, one can eliminate $x$ and $z$ components of $\boldsymbol{t}_{ij}^{\phantom{0}}$ in the nearest \texttt{nc}-bonds between the $ac$ planes (Fig.~\ref{fig.CuO}d). The remaining $D_{ij}^{y}$ components do not contribute to the spiral magnetic order, and the latter emerges solely from the competition of isotropic $J_{ij}$'s~\cite{SM}. The same holds for $\vec{\boldsymbol{\mathcal{A}}}_{ij} = [ \mathcal{A}_{ij}^{v,c} ]$, whose only non-zero component is $c = y$, and there is no contribution to $\vec{P}$. The isotropic $\vec{\mathsf{P}}_{ij}$ in the \texttt{nc}-bonds can also be large~\cite{SM}, but does not add up to $\vec{P}$ because of the $90^{\circ}$ alignment of spins.
\end{thebibliography}
\end{document}